\documentclass{aa}  

\usepackage{psfig}
\usepackage{natbib}
\usepackage[english]{babel}  
\usepackage{graphicx}
\usepackage{txfonts}
\begin{document}

   \title{\emph{XMM-Newton} observation of the supernova remnant Kes 78 (G32.8-0.1): Evidence for shock-cloud interaction}

   \author{M. Miceli
          \inst{1,2}
          \and
          A. Bamba\inst{3,4}
          \and
          S. Orlando\inst{2}
          \and
          P. Zhou\inst{5}
          \and
          S. Safi-Harb\inst{6}
          \and
          Y. Chen\inst{5}
          \and
          F. Bocchino\inst{2}
          }

   \institute{Dipartimento di Fisica \& Chimica, Universit\`a di Palermo, Piazza del Parlamento 1, 90134 Palermo, Italy\\
              \email{miceli@astropa.unipa.it}
         \and
     INAF-Osservatorio Astronomico di Palermo, Piazza del Parlamento 1, 90134 Palermo, Italy
         \and    
     Department of Physics, Graduate School of Science, The University of Tokyo, 7-3-1 Hongo, Bunkyo-ku, Tokyo 113-0033, Japan
     \and
     Research Center for the Early Universe, School of Science, The University of Tokyo, 7-3-1 Hongo, Bunkyo-ku, Tokyo 113-0033, Japan
     \and
     Department of Astronomy, Nanjing University, Nanjing 210023, China
     \and
     Department of Physics and Astronomy, University of Manitoba, Winnipeg, MB R3T 2N2, Canada
     }

   \date{}

 
  \abstract
   {The Galactic supernova remnant Kes 78 is surrounded by dense molecular clouds, whose projected position overlaps with the extended HESS $\gamma-$ray source HESS J1852-000. The X-ray emission from the remnant has been recently revealed by \emph{Suzaku} observations, which have shown indications for a hard X-ray component in the spectra, possibly associated with synchrotron radiation.}
   {We aim at describing the spatial distribution of the physical properties of the X-ray emitting plasma and at revealing the effects of the interaction of the remnant with the inhomogeneous ambient medium. We also aim at investigating the origin of the $\gamma-$ray emission, which may be Inverse Compton radiation associated with X-ray synchrotron emitting electrons or hadronic emission originating from the impact of high energy protons on the nearby clouds.}
   {We analyzed an \emph{XMM-Newton} EPIC observation of Kes 78 by performing image analysis and spatially resolved spectral analysis on a set of three regions. We tested our findings against the observations of the $^{12}$CO and $^{13}$CO emission in the environment of the remnant.}
   {We revealed the complex X-ray morphology of Kes 78 and found variations of the spectral properties of the plasma, with significantly denser and cooler material at the eastern edge of the remnant, which we interpret as a signature of interaction with a molecular cloud. We also exclude the presence of narrow filaments emitting X-ray synchrotron radiation.}
   {Assuming that the very high energy $\gamma-$ray emission is associated with Kes 78, the lack of synchrotron emission rules out a leptonic origin. A hadronic origin is further supported by evidence for interaction of the remnant with a dense molecular cloud in its eastern limb.}

 
\titlerunning{Evidence for shock-cloud interaction in Kes 78}
   \maketitle
%

\section{Introduction}

Supernova remnants (SNRs) are considered to be the main sources of Galactic cosmic rays (CRs) up to the knee of the CR energy distribution ($\sim3$ PeV).
The current generation of TeV and GeV observatories has allowed us to reveal the $\gamma$-ray emission of a rapidly growing number of SNRs. This emission can be a signature of proton acceleration, if associated with $\pi^0$ decay. In this scenario, the $\pi^0$ decays follow proton-proton collisions due to accelerated hadrons impacting the ambient medium (this is known as the hadronic scenario). However, the $\gamma$-ray emission can also be the result of inverse Compton (IC) scattering from ultrarelativistc electrons (accelerated at the shock front) on the cosmic microwave background and/or dust-emitted infrared photons (leptonic scenario, see \citealt{rey08} for further details). 

It is typically difficult to ascertain the origin of the $\gamma$-ray emission in young, X-ray synchrotron emitting SNRs, such as Vela Jr. (\citealt{tab11}); RX J1713.7-3946 (\citealt{eps10,esp12,ga14,zc16}); and SN 1006 (\citealt{aaa11,alr15,mad14,mop16}). However, $Fermi$ and $AGILE$ observations of middle-aged SNRs, characterized by thermal X-ray emission and lack of synchrotron X-ray radiation, clearly revealed the presence of high energy hadrons, e.~g.,  W44 \citep{aaaas10}, W28 \citep{aaaa10}, and IC 443 \citep{tgc10}. All these SNRs interact with dense clouds and the hadronic emission seems to originate from CRs leaving the acceleration site and interacting with the dense ambient material.

The Galactic supernova remnant Kes 78 (G32.8-0.1) shows different signatures of interaction with ambient clouds. In particular, the OH masers detected by \citet{gfg97} reveal shock-excited emission due to the interaction of the SNR with an adjacent molecular cloud \citep{kfg98}, and the $^{12}$CO emission exhibits a cavity structure, which is consistent with the radio contours of the remnant (\citealt{zzz07,zc11}). Also, the analysis of the $^{13}$CO and $^{12}$CO line emission performed by \citet{zc11} strongly suggests an interaction with a large cloud in the east. The position of this eastern molecular cloud 
coincides with that of the $\gamma-$ray extended source HESS J1852-000 (\citealt{kca11})\footnote{A $Fermi-LAT$ GeV source (2FGL J1850.7-0014c, \citealt{aaa15}) was found near Kes 78, but the significance of the detection is under debate (\citealt{asc14}).}.

Recently, \citealt{bth16} (hereafter B16) revealed for the first time the complex morphology of the X-ray emitting plasma in Kes 78 by analyzing dedicated \emph{Suzaku} observations (previously, only the X-ray emission from a small region in the northern edge of the shell had been studied by \citealt{zc11}). The X-ray emission presents a complex morphology, well correlated with the radio emission, visible in the $0.5-2~$keV energy band. Below 2.5 keV, the \emph{Suzaku} spectra can be well modelled by a soft thermal component associated with an underionized plasma with temperature $kT\sim0.6-0.7$ keV and with a very low density $n\sim 10^{-3}-10^{-2}$ cm$^{-3}$, consistent with a low density cavity. B16 also noticed a fainter hard X-ray component which dominates the spectrum at energies $>2.5$ keV. Though the morphology of this component is not revealed in the hard X-ray maps, spectra extracted from the northern and southern parts of the remnant both exhibit this high energy tail, which may be thermal emission from a very hot plasma ($kT=3.4^{+1.3}_{-1.2}$ keV) or nonthermal radiation with a photon index $\Gamma=2.3\pm0.3$. The latter scenario may indicate that high energy electrons are accelerated at the shock front of Kes 78 and produce synchrotron radiation. These high-energy electrons may then up-scatter the cosmic microwave background photons up to the $HESS$ energy band, thus explaining the TeV detection in terms of leptonic emission. On the other hand, B16 note that the luminosity of the hard component is a few orders of magnitude lower than that of other synchrotron emitting SNRs  and also the shock velocity estimated from the remnant size and age ($v_{s} \sim 5-7 \times10^{7}$ cm s$^{-1}$) is slower than the typical value (a few 10$^8$ cm s$^{-1}$) observed in shocks emitting synchrotron X-rays. 

We here present the analysis of an \emph{XMM-Newton} EPIC observation of Kes 78. We aim at: i) studying the spatial distribution and the physical origin of the soft$/$hard X-ray emission in the remnant, ii) investigating the shock-cloud interaction process, and iii) deriving information on the origin of the $\gamma-$ray emission.

The paper is organized as follows: the data and the data analysis procedures are presented in Sect. \ref{data}, the results of the image analysis and of the spatially resolved spectral analysis are shown in Sect. \ref{Image analysis} and Sect. \ref{Spectral analysis}, respectively. Finally, our conclusions are discussed in Sect. \ref{Conclusions}

\section{Data Analysis}
\label{data}

We analyzed the \emph{XMM-Newton} EPIC observation ID 0671510101 (PI F. Acero) performed in March 2012 with pointing coordinates $\alpha_{J2000}=18^h51^m33.5^s$, $\delta_{J2000}=-00^{\circ}10'10.0"$. The pn camera operated in Full Frame Mode and the MOS cameras in Large Window mode (all the cameras with the medium filter). 
Data were processed with the Science Analysis System (SAS V15). We selected events with PATTERN$\le12$ for the MOS cameras, PATTERN$\le$4 for the pn camera, and FLAG=0 for both. We then adopted the ESPFILT task (based on a sigma-clipping algorithm) to remove high background periods from the event lists, thus obtaining a screened exposure time of $27,~36,~38$ ks for the pn,~MOS1, and MOS2 observations, respectively. We used the task edetect\_chain (with detection likelihood threshold set to 10) in the $2.5-6$ keV energy band to detect hard point-like sources.

All the images presented here are superpositions of the MOS1, MOS2, and pn images (obtained using the $EMOSAIC$ task) and are vignetting-corrected and adaptively smoothed by adopting the procedure described in \citet{mdb06} (see their Sect. 2). All images are background-subtracted through a double subtraction procedure, as described below. We first considered the EPIC high signal-to-noise background event files (hereafter bkg files) m1\_m\_ff\_g, m2\_m\_ff\_g, pn\_m\_ef\_g, (\citealt{cr07}). We then estimated and subtracted the quiescent non-photonic background in our data and in the bkg files by scaling the EPIC Filter Wheel Closed (FWC) observations\footnote{For the repository of the EPIC background and FWC event files, see \url{http://xmm-tools.cosmos.esa.int/external/xmm_calibration/background/bs_repository/blanksky_all.html}, and \url{http://www.cosmos.esa.int/web/xmm-newton/filter-closed}, respectively.}, using, as a scaling factor, the ratio of the count rates in the unexposed corners of the detectors before subtracting. We finally subtracted the ``pure photonic" background to our pure photonic data.

We also present a median photon energy map (hereafter $E_m$ map) which is an image where each pixel holds the median energy of the pn photons detected therein in the $0.5-4$ keV energy band. This map provides information about the spatial distribution of the spectral hardness of the X-ray emission. The map has a bin size 18$''$ so as to collect more than 4 counts per pixel and more than 12 counts in all the pixels within the Kes 78 shell. The map is smoothed by adopting the procedure described in \citet{mdb08}, with $\sigma=72''$.

Spectral analysis was performed in the $0.5-4.0$ keV energy band using XSPEC v12.9 \citep{arn96}. We processed the event files with the EVIGWEIGHT task \citep{ana01} to correct for vignetting effects. Spectra were extracted with the EVSELECT task and the response files were generated with the ARFGEN and RMFGEN tasks. In the spatially resolved spectral analysis (discussed in Sect. \ref{Spectral analysis}, see Fig. 1 for the extraction regions) we rebinned all the spectra to achieve a signal-to-noise ratio per bin $>5\sigma$. To obtain tight constraints on the hard X-ray spectrum of the whole remnant and to reduce the fluctuations associated with background subtraction, we rebinned the global spectrum to achieve a signal-to-noise ratio per bin $>10\sigma$. For each spectrum, we subtracted a background spectrum extracted from a nearby region immediately outside of the SNR shell. We verified that the best-fit values do not depend significantly on the choice of the region and all the best-fit parameters are consistent within less than one sigma. We then show only the results obtained by adopting the background region shown in the left panel of Fig. 1 (dashed white ellipse). MOS and pn spectra of the different observations were fitted simultaneously. To describe the thermal emission, we adopted the VNEI model which describes an isothermal optically thin plasma in non-equilibrium ionization and is based on AtomDB V3.0. The interstellar absorption is calculated through the TBABS model.
All the reported errors are at 90\% confidence, as calculated according to \citet{lmb76}.

\section{Results}
\label{Results}
 
We first performed an image analysis of the EPIC data to study the morphology of the remnant and to identify interesting regions which we then investigated with a spatially resolved spectral analysis.

\subsection{Image analysis} 
\label{Image analysis} 

\begin{figure*}[htb!]
 \centerline{\hbox{\psfig{figure=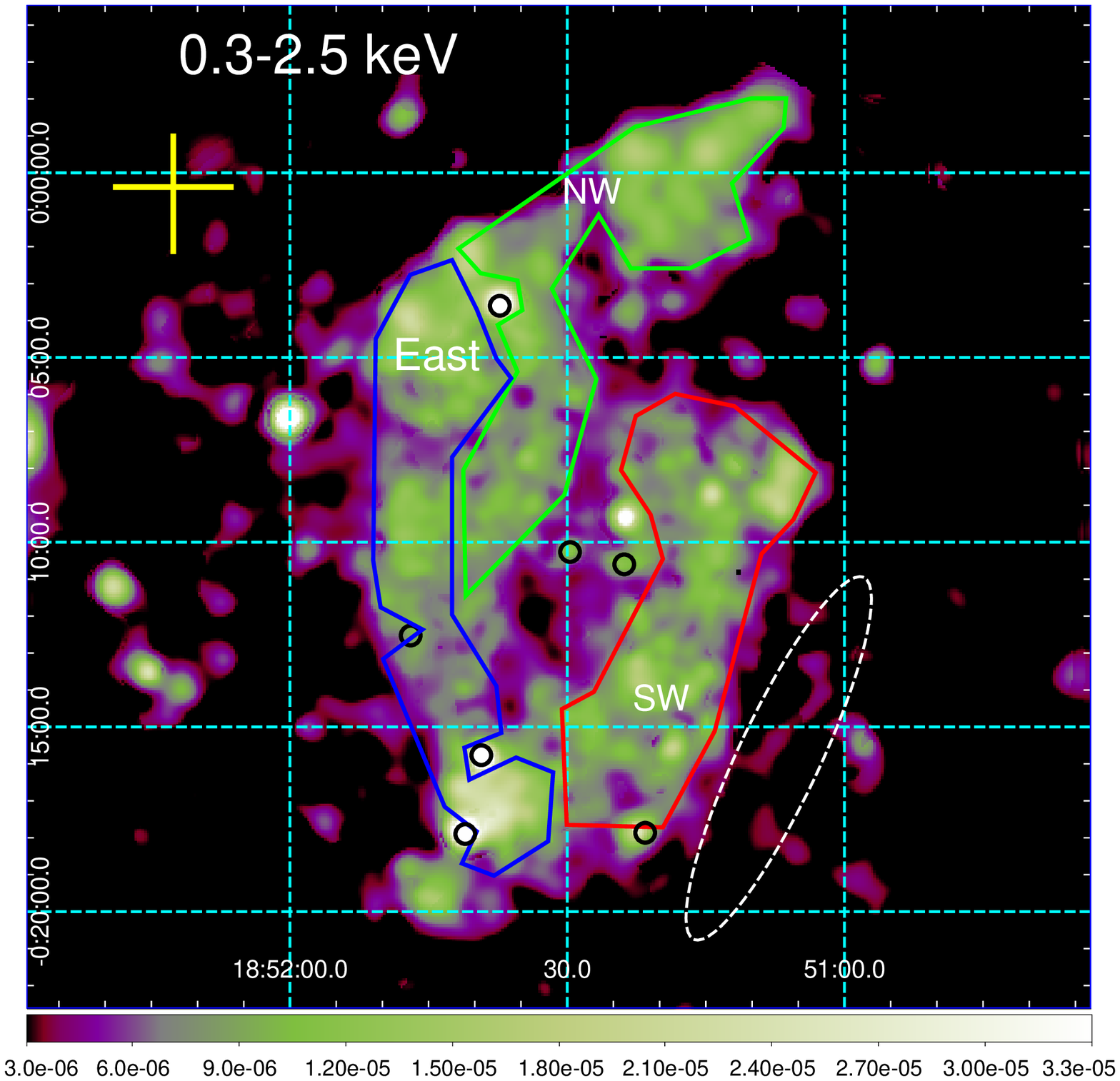,width=\columnwidth}}
 \hbox{\psfig{figure=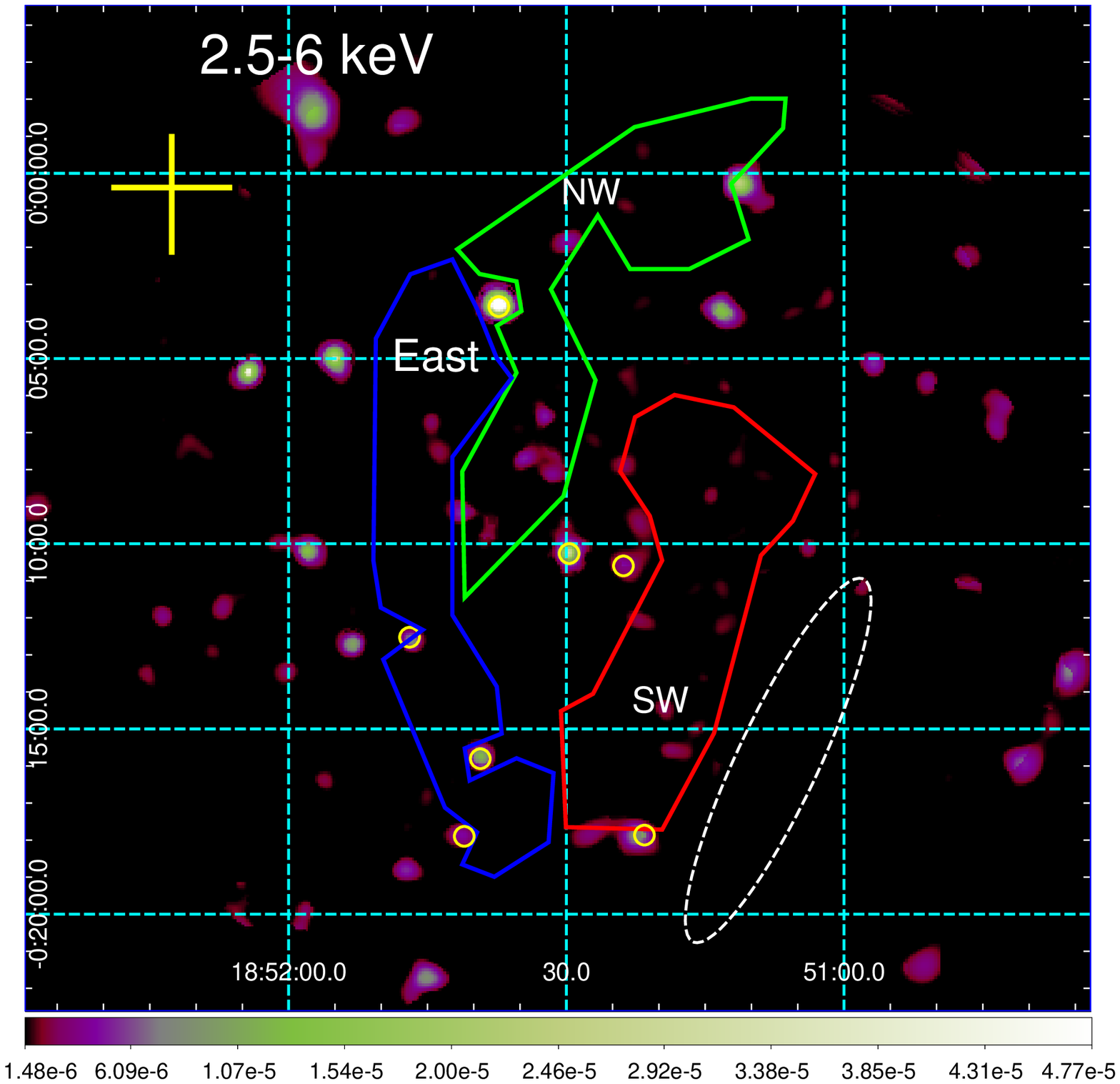,width=\columnwidth}}}

\caption{\emph{Left panel:} EPIC count-rate images (MOS and pn mosaic) of Kes 78 in the $0.3-2.5$ keV band (counts per second per $4''$ bin). The image is background subtracted, vignetting-corrected and adaptively smoothed to a signal-to-noise ratio of 16. North is up and East is to the left. Regions East, Northwest, and Southwest (selected for the spatially resolved spectral analysis) are shown in blue, green, and red, respectively. The region selected to extract the background spectrum is indicated by the dashed white ellipse. Hard point-like sources are indicated by black circles. The yellow cross to the north-east shows the centroid of the extended $\gamma-$ray source HESS J1852-000 \citep{kca11}.  \emph{Right panel:} same as left panel in the $2.5-6$ keV band (with hard point sources shown in yellow).}
\label{fig:Xmaps}
\end{figure*}

Figure \ref{fig:Xmaps} shows the EPIC count-rate map of Kes 78 in soft ($0.3-2.5$ keV, left panel) and hard ($2.5-6~$keV, right panel) X-rays. The soft map shows that the overall morphology of the remnant is far from being spherical, as already shown by the \emph{Suzaku} observations analyzed by B16. The higher spatial resolution of the \emph{XMM-Newton} EPIC data allows us to reveal new details of the very complex morphology, which is characterized by two arc-shaped structure: one running from south to northwest (corresponding to regions ``East" and ``NW" in the figure) and the other, with a different concavity, running in the southeast-northwest direction (region ``SW" in the figure). 

The background-subtracted count rate-image in the $2.5-6$~keV energy band (right panel of Fig. \ref{fig:Xmaps}) reveals a set of hard point-like sources, but does not show any diffuse emission. This is similar to what B16 found in the \emph{Suzaku} images (the hard X-ray component was detected in the \emph{Suzaku} spectra, but not imaged in the maps). However, if the hard X-ray emission observed by B16 were indeed synchrotron emission, we would have expected to observe fine structures in our hard EPIC X-ray maps. In fact, synchrotron emission should be localized in narrow filaments (at the shock front), which are typically smaller than the point spread function of the \emph{Suzaku} telescope, but very well resolved with \emph{XMM-Newton} (\citealt{byu03,bal06,mbd13}). 
On the other hand, the \emph{XMM-Newton} instrumental background in the hard X-rays is higher than that of \emph{Suzaku} and the hard X-ray emitting component detected by B16 is relatively faint and may be distributed over a large area (especially if it has a thermal origin). Therefore, the hard X-ray diffuse emission, if any, may be below the background. We will provide quantitative upper limits on this hard X-ray component in Sect. \ref{Spectral analysis}.
\begin{figure}[htb]
  \centerline{\hbox{\psfig{figure=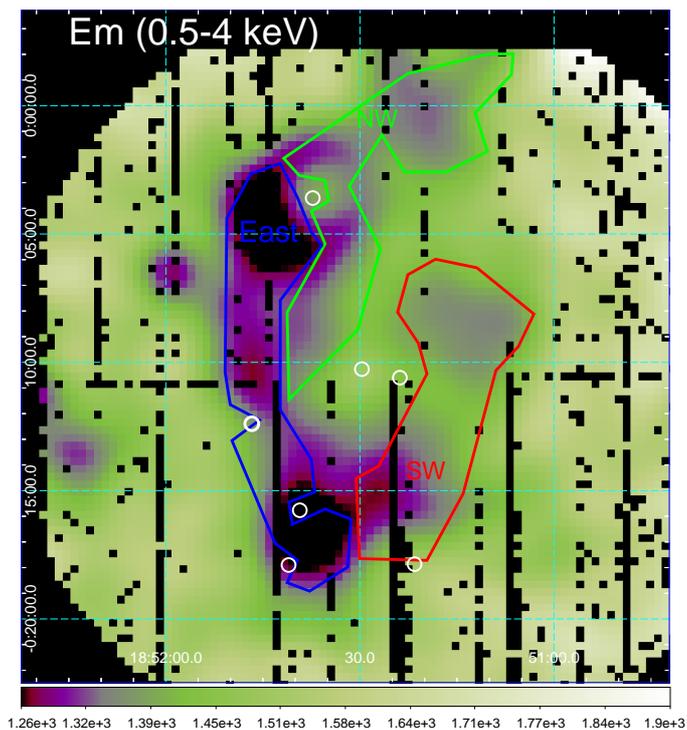,width=\columnwidth}}}
\caption{Median photon energy ($E_m$) map of Kes 78. Each pixel stores the local median photon energy (in eV) of the photons detected by the pn camera in the $0.5-4$ keV energy band. The bin size is $18''$ and the map is smoothed by adopting a Gaussian kernel with $\sigma=72''$. Pixels with less than 4 counts have been masked out. White circles show the hard point-like sources within the Kes 78 shell.}
\label{fig:MPE1}
\end{figure}

\begin{figure}[h]
  \centerline{\hbox{\psfig{figure=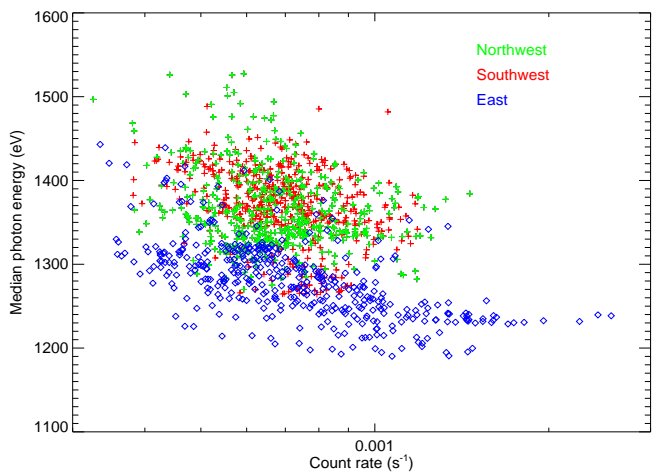,angle=90,width=\columnwidth}}}
\caption{$E_m$ vs. count-rate scatter plot for the pn pixels in the regions shown in Fig. \ref{fig:MPE1}. The blue, green, and red points corresponds to pixels in regions East, Northwest, and Southwest, respectively.}
\label{fig:MPE2}
\end{figure}

Figure \ref{fig:MPE1} shows the median photon energy map of Kes 78 in the $0.5-4$ keV energy band. We observe significant inhomogeneities in the $E_m$, with region East showing lower values than regions Northwest and Southwest. This result strongly suggests the presence of inhomogeneities in the plasma temperature, with cooler plasma producing softer thermal X-rays in the eastern limb. In the case of pressure equilibrium, we expect the cooler plasma to be denser than the hotter plasma, thus having a higher emission measure. The radiative losses function per unit emission measure (\citealt{rs77,mgv85}) increases with lower temperatures, for temperatures in  the range $10^{6}-10^{7}$ K (i.e., those measured in Kes 78, see Sect. \ref{Spectral analysis} and B16). Therefore, if the inhomogeneities in the median photon energy originate from different thermal conditions in the plasma, we expect higher count-rates in the soft X-ray emitting regions (i.e., region East). Another possibility is that the $E_m$ fluctuations are not intrinsic, but are instead the result of inhomogeneities in the absorbing column density: higher values of $N_H$ in region Northwest and Southwest may, in fact, absorb efficiently the soft X-ray emission thus explaining the observed harder X-ray emission therein. In this case we would expect lower count-rates in the regions with low $E_m$.

To discriminate among these two scenarios, we produced the $E_m$ vs. count-rate scatter plot shown in Fig. \ref{fig:MPE2}, where each point reveals the values of count-rate and $E_m$ (in the $0.5-4$ keV band) measured in the pixels located inside region East (shown in blue), Northwest (green), and Southwest (red). We find that the global trend is a decreasing one, as expected in the case of pressure equilibrium (e.g., \citealt{bmm04,mbm05}). Also, we reveal two different regimes with the points of region East clearly identifying a narrow strip, characterized by low $E_m$ and reaching high count-rates, while the points of both region Northwest and Southwest show higher $E_m$ and, on average, lower count-rates. This is the trend expected when $E_m$ variations originate with differences in the plasma temperature. We then conclude that we expect the plasma to be cooler in the eastern limb of Kes 78, the temperature being higher and relatively uniform elsewhere. 

\subsection{Spectral analysis} 
\label{Spectral analysis} 

\begin{figure}[htb!]
 \centerline{\hbox{\psfig{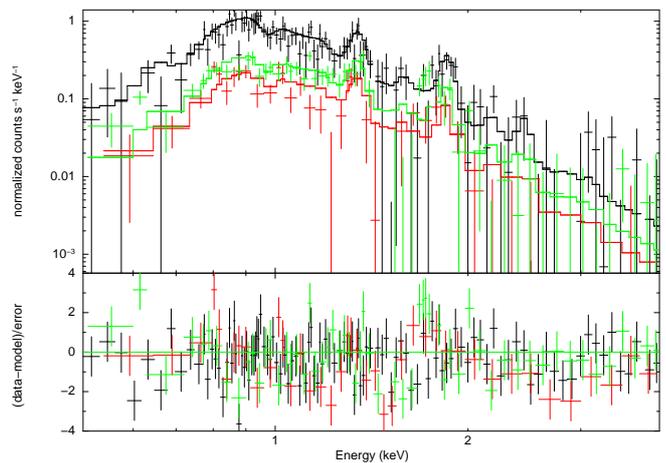}}}
\caption{\emph{Upper panel:} EPIC pn (black) and MOS1,2 (red, green) global spectra of Kes 78 together with their best fit model and residuals (displayed in the lower inset).  }
\label{fig:globspec}
\end{figure}

\begin{figure}[htb]
 \centerline{\hbox{\psfig{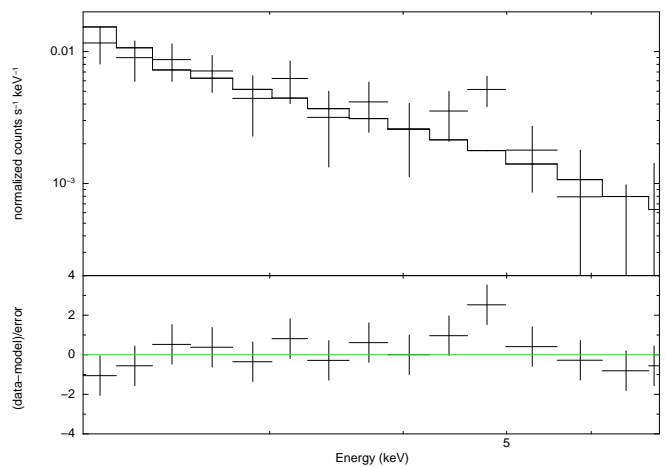}}}
\caption{\emph{Upper panel:} EPIC pn cumulative spectrum of the hard X-ray point sources within the shell of Kes 78 together with the (properly scaled) best-fit model adopted by B16 for the hard X-ray emitting component and corresponding residuals (see Sect. \ref{Spectral analysis} for details).  }
\label{fig:ptsrc}
\end{figure}

\begin{figure}[ht!]
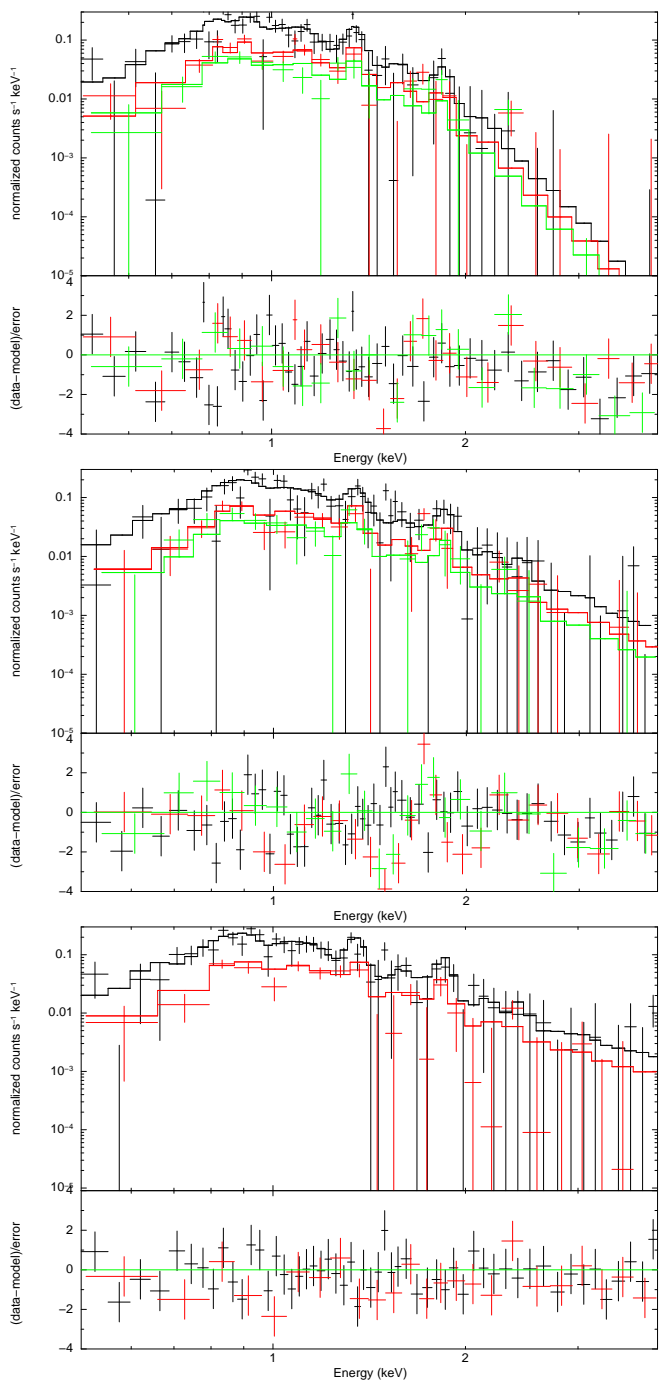

 \centerline{\hbox{\psfig{figure=specEast.ps,angle=-90,width=\columnwidth}}}
 \centerline{\hbox{\psfig{figure=specSW2.ps,angle=-90,width=\columnwidth}}}
\centerline{\hbox{\psfig{figure=specNW.ps,angle=-90,width=\columnwidth}}}
 \caption{\emph{Upper panel:} EPIC pn (black) and MOS1,2 (red, green) spectra extracted from region East of Fig. \ref{fig:Xmaps} together with their best fit model and residuals (displayed in the lower inset).  \emph{Central panel:} same as upper panel for region Southwest. \emph{Lower panel:} same as upper panel for region Northwest, MOS1 spectrum is not shown because the extraction region falls in a damaged CCD chip.}
\label{fig:spec}
\end{figure}

We first studied the global spectra of the remnant, by analyzing the spectra extracted from the union of the three regions shown in Fig. \ref{fig:Xmaps}. We found that a single isothermal component of an optically thin plasma in non-equilibrium ionization provides a good fit of the EPIC spectra (reduced $\chi^2=1.30$ with 517 d.o.f.). In the fittings we first let the Fe, Ne, Mg, and Si (i.e., those elements whose line complexes are clearly visible in the spectra) abundances free to vary. We found that the best-fit values of the Fe, Ne, and Si abundances are all consistent with being solar and that the improvement in the fits is not significant. We then fixed all these abundances to their solar values, by letting only the Mg free to vary. The global X-ray spectrum of Kes 78 together with its best-fit model is shown in Fig. \ref{fig:globspec} and the best-fit parameters are given in Table \ref{tab:fit} and are consistent (within two sigmas) with those obtained with \emph{Suzaku} by B16 (for their soft component). 
\begin{table*}[htb!]
\caption{Best-fit parameters of spectral fittings of the global spectrum of Kes 78 and of the regions shown in Fig. \ref{fig:Xmaps}}
\centering                        
\begin{tabular}{c c c c c} 
\hline\hline               
Parameters                  &    Global    &        East       &  Southwest   & Northwest \\    
\hline                        
$N_H$ ($10^{21}$ cm$^{-2}$)&  $8.3\pm0.7$  & $13.1^{+0.9}_{-1.4}$ &$8.8^{+1.5}_{-1.2}$ & $8.3^{+1.7}_{-1.3}$ \\   
  $kT$ (keV)          & $0.8^{+0.4}_{-0.2}$  & $0.26^{+0.10}_{-0.05}$ &$0.9^{+0.7}_{-0.3}$ & $1.8^{+2}_{-1.0}$ \\
 Mg$/$Mg$_\sun$       & $1.2\pm0.3$ & $0.9\pm0.2$ & $1.3\pm0.4$ & $0.9\pm0.2$ \\
$\tau$ ($10^{10}$ cm$^{-3}$ s)  & $4^{+3}_{-2}$  & $10^{+12}_{-7}$ & $4^{+5}_{-2}$ & $2^{+3}_{-1}$  \\
$EM^{*}$ ($10^{18}$ cm$^{-5}$) & $0.19^{+0.09}_{-0.06}$ & $6^{+10}_{-4}$    &$0.15^{+0.15}_{-0.07}$ & $0.1^{+0.1}_{-0.05}$ \\ 
reduced $\chi^2$ (d.o.f.) & $1.30~(517)$  &  $1.17~(518)$    &$1.15~(569)$ & $1.05~(466)$ \\
\hline                                   
\end{tabular}
\tablefoot{$^*$ Emission measure per unit area.}
\label{tab:fit}
\end{table*}

We then investigated the presence of a hard X-ray emitting component in the EPIC spectra and verified that this additional component is $not$ statistically needed. We added to the best fit model reported in Table \ref{tab:fit} either a power law component or a hotter thermal component. In both cases, there are no improvements in the fits and the normalization of the additional component is consistent with zero (at one sigma). Both image and spectral analysis concur in excluding a diffuse hard X-ray emission in Kes 78. We already noted in Sect. \ref{Image analysis} that this may be because of the relatively high \emph{XMM-Newton} background. We here investigate this issue in detail. In the global spectrum of Kes 78, the flux above 2 keV is dominated by the background and is $F_h= 6\pm1\times10^{-12}~$erg$~$cm$^{-2}$$~$s$^{-1}$ which is much higher than the flux of the hard X-ray component observed by B16, which is $\sim3.5\pm0.7\times10^{-13}$ erg cm$^{-2}$ s$^{-1}$ (though this value was measured in a slightly smaller region, given that the \emph{Suzaku} observation do not cover the eastern edge of the remnant). Therefore, if the hard X-ray emission were smoothly distributed over the whole shell, it would not be possible to detect it with our \emph{XMM-Newton} observations. 

On the other hand, if we assume that the hard X-ray emission has a nonthermal origin, we expect it to be concentrated in narrow rims at the border of the shell, marking the position of the shock front. We then considered a narrow region, $16''$ wide, running at the intersection between the border of the shell and the spectral region adopted by B16. In this region the flux of the background in the $2-10$ keV band is about $75\%$ as that of the hard component observed by B16. Therefore, if this component were present, the hard X-ray surface brightness would be larger by more than a factor of two at the border of the shell than elsewhere. This is at odds with what we found in our X-ray maps (see right panel of Fig. \ref{fig:Xmaps}). Therefore, the hard X-ray emission of Kes 78 is not concentrated in narrow filaments. There are indeed a few cases of middle-aged SNRs, namely RX J1713.7-3946 \citep{abd09} and (part of) RCW 86 \citep{bcv14}, whose synchrotron emission appears to be distributed over wide regions and not concentrated in narrow filaments. However, both these remnants have a hard X-ray luminosity which is a factor of $40-60$ times higher than that estimated by B16 for Kes 78 (see \citealt{nbd12}).

We also note that the regions selected by B16 include 6 hard X-ray emitting point-like sources (shown in Fig. 1), which are unresolved in the \emph{Suzaku} data. We extracted the cumulative spectrum of all these sources and found that it can be fitted by the same model as that adopted by the B16 for the hard component, though with a lower hard X-ray flux, which is $F_{pt}=1.2\pm0.3\times10^{-13}$ erg cm$^{-2}$ s$^{-1}$. The spectrum with its model (absorbed power law with photon index $\Gamma=2.3$, as in B16) is shown in Fig. \ref{fig:ptsrc}. It is possible that these sources have contaminated the \emph{Suzaku} spectrum of Kes 78 by artificially enhancing its hard X-ray emission. It is difficult to quantify this contamination given that we do not know whether these sources are variable or not. Also, 4 of these 6 sources are at (or close to) the border of the regions analyzed by B16. Therefore, considering the high \emph{Suzaku} PSF, part of their flux has possibly been scattered out of the regions by the \emph{Suzaku} mirrors. On the other hand, the hard X-ray photons from other point-like sources outside the region may have been scattered inside, especially considering that the variable source 3XMM J185114.3-000002 (see Sect. \ref{3XMM}) was in a very high state during the \emph{Suzaku} observation. However, if we subtract $F_{pt}$ from $F_h$, we still have a hard X-ray flux that, if present, would have been detected in our EPIC data, if concentrated in narrow rims at the shock front.

We then conclude that any hard X-ray emission of Kes~78 cannot be associated with synchrotron radiation from high-energy electrons accelerated at the shock front.

To further investigate the spatial distribution of the thermal properties of the plasma, we then performed a spatially resolved spectral analysis by extracting the EPIC spectra from the three spatial regions shown in Fig. \ref{fig:Xmaps}. The study of the $E_m$ performed in Sect. \ref{Image analysis} shows that region East is characterized by low values of $E_m$ and high count rates, while regions Northwest and Southwest have harder X-ray emission.  

Figure \ref{fig:spec} shows the spectra extracted from regions East, Southwest, and Northwest.
In all the regions, we obtained a good fit of the X-ray spectra by adopting a single isothermal component and the best-fit parameters are reported in Table \ref{tab:fit}. The best-fit values obtained in region Northwest are in good agreement with those derived for a similar region studied by \citet{zc11}.
We found that the plasma temperature, $kT$ is lower in region East than in regions Northwest and Southwest, in agreement with the prediction of our image analysis.  Also, the plasma emission measure per unit area ($EM$, which is the product of the square of the plasma density, $n$ times its extension along the line of sight, $\Lambda$) is higher in region East than elsewhere.

\begin{figure}[htb!]
 \centerline{\hbox{\psfig{figure=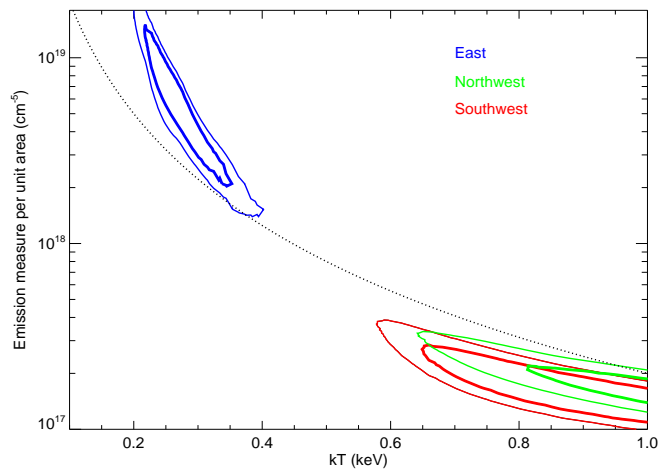,angle=90,width=\columnwidth}}}
\caption{The $68\%$ (thick lines) and $90\%$ (thin lines) confidence contour levels of the plasma emission measure per unit area, $EM$, versus the temperature, $kT$, for regions East (blue), Northwest (green) and Southwest (red). The dotted curve shows the expected trend for increasing ambient density ($EM\propto 1/(kT)^2$), assuming the same extension along the line of sight in all the regions (see Sect. \ref{Conclusions}).}
\label{fig:steppar}
\end{figure}

Figure \ref{fig:steppar} shows the confidence contour levels of $EM$, versus $kT$ for the three regions revealing the significance of these inhomogeneities. While regions Northwest and Southwest have consistent values of $EM$ and $kT$, these values are significantly different in region East, where $kT$ is lower by a factor $f_{kT}\sim 2-7$ and $EM$ is higher by a factor $f_{EM}\sim10-100$, corresponding to an increase by a factor $f_n\sim3-10$ in the plasma density, by assuming the same value of $\Lambda$ for the three regions\footnote{Different extensions along the line of sight between two regions would introduce a factor $l^{-1/2}$, where $l$ is the ratio between the corresponding values of $\Lambda$.}.

We also found variations in the absorbing column density, $N_H$, which is higher in region East than in regions Northwest and Southwest. This result further confirms that the variations in the $E_m$ map are due to inhomogeneities in the plasma temperature and not in the interstellar absorption, because in this case we would have expected a lower value of $N_H$ in region East. As a test, we imposed the $N_H$ in region East to be the same as that in region Southwest and checked how this affect the quality of the fits and the inhomogeneities in the plasma $kT$ and $EM$. We found that the reduced $\chi^2$ is higher with the lower $N_H$, being $1.20$ with 519 d.o.f., to be compared with $1.17$ with 518 d.o.f. (see Table \ref{tab:fit}), and The F-test gives the probability that the improvement of the fit obtained with a higher $N_H$ is insignificant as $0.1\%$. Moreover, even with this value of $N_H$, we still obtain a lower best-fit temperature in region East, with $kT=0.57\pm0.08$ keV.
We then conclude that the physical conditions of the X-ray emitting plasma in region East are significantly different from other regions.

We obtained an independent evaluation of the gas density in our spectral regions by analyzing the Kes 78 maps in the $^{13}$CO line emission (first presented in \citealt{zc11}). 
We estimated the $^{13}$CO column density with the Gaussian components of the $^{13}$CO emission around the systemic velocity of $81$ km s$^{-1}$, in order to avoid contamination from irrelevant molecular lines in the line of sight. The $^{13}$CO emission was assumed to be optically thin and in local thermodynamic equilibrium. Taking an abundance ratio H$_2$$/$$^{13}$CO$=$$5\times 10^5$ \citep{dic78} and a kinetic temperature of 10 K for typical interstellar molecular clouds, we derived the column densities of H$_2$ ($N_{H_2}$) in region East ($N_{H_2}\sim 2.8\times 10^{21}$ cm$^{-2}$) and Southwest ($N_{H_2}\sim0.9\times10^{21}$  cm$^{-2}$).  We then found that $N_{H_2}$ in region East is expected to be higher by $\Delta N_{H_2}\sim 2\times 10^{21}$ cm$^{-2}$ than in region Southwest. This is in qualitative agreement with the findings of the X-ray data analysis, which suggest $\Delta N_H=1.4-6.4\times10^{21}$ cm$^{-2}$. 

\begin{figure*}[htb!]
 \centerline{\hbox{\psfig{figure=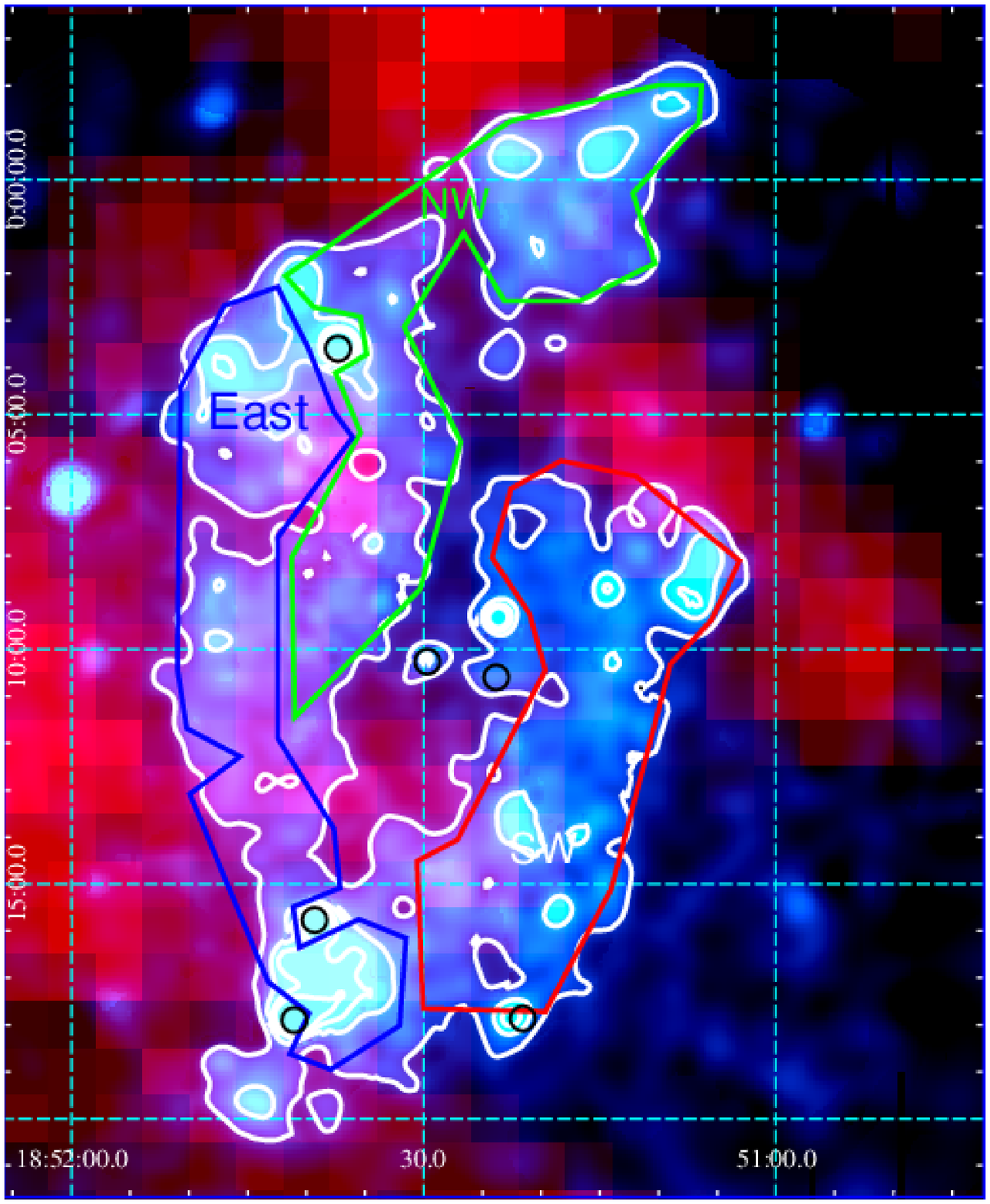,width=\columnwidth}}
\hbox{\psfig{figure=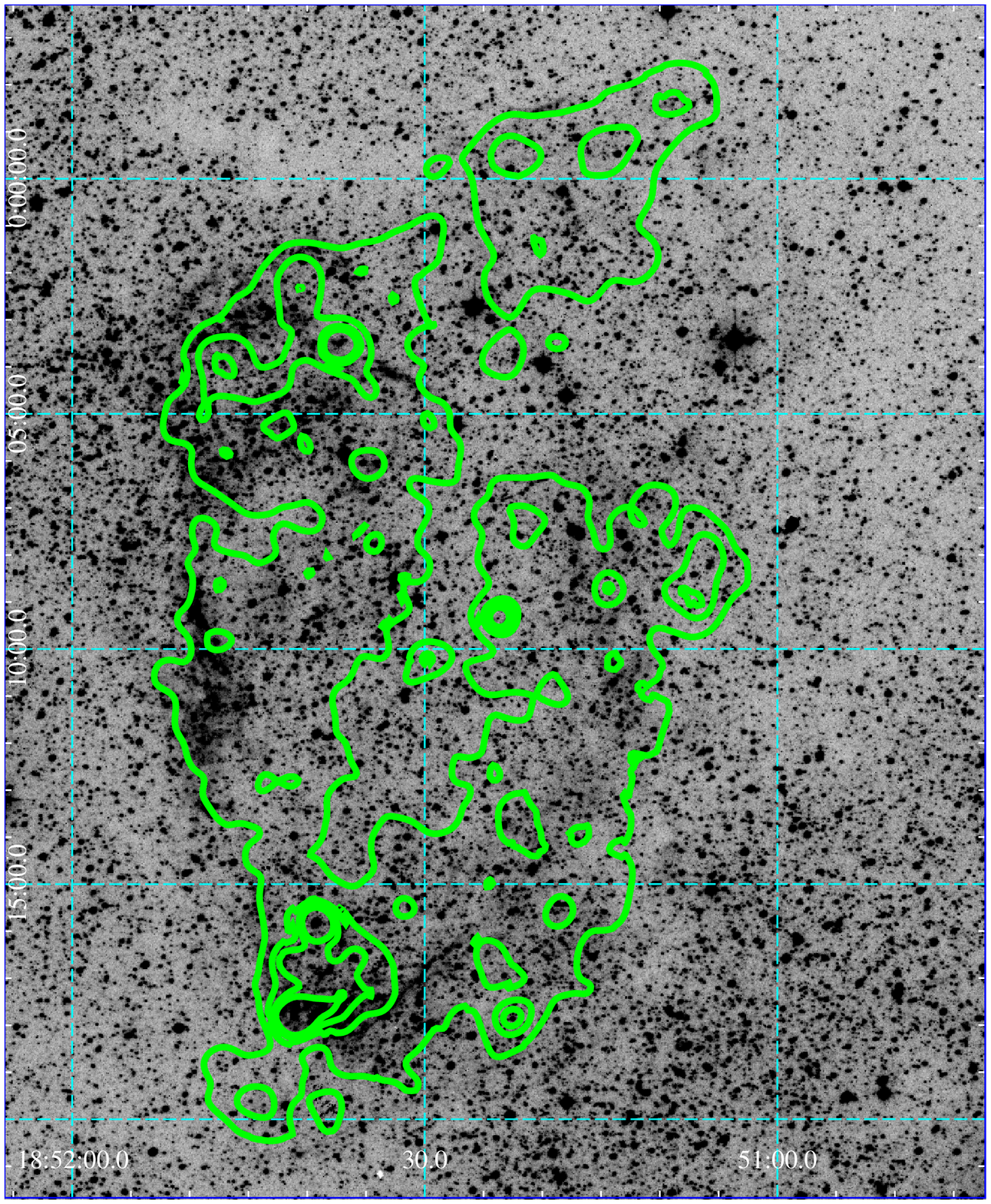,width=\columnwidth}}}
\caption{\emph{Left panel:} PMOD $^{12}$CO (J$=1-0$) image in the $80-84$ km s$^{-1}$ velocity range (red) together with the \emph{XMM-Newton} EPIC observation of Kes 78 in the $0.3-2.5$ keV band (light blue). The white curves are the X-ray contours levels \emph{Right panel:} H$_\alpha$ image of the Kes 78 environment obtained within the AAO/UKST SuperCOSMOS H$_\alpha$ survey with the X-ray contours superimposed in green.}
\label{fig:COHa}
\end{figure*}

\subsection{The variable source 3XMM J185114.3-000002}
\label{3XMM}

B16 analyzed the X-ray emission from the variable source 2XMM J185114.3-000004, which is also catalogued in the 3XMM Data Release 5 \citep{rww16} as  3XMM J185114.3-000002, with coordinates $\alpha_{J2000}=18^h51^m14.1^s$, $\delta_{J2000}=-00^{\circ}00'02.8"$.
B16 found the source to be in a very high state with a flux $F\sim10^{-11}$ erg cm$^{-2}$ s$^{-1}$ and an intrinsic luminosity $L_X\sim1.1\times10^{35}$ erg s$^{-1}$ in the $2-10$ keV band. They also detected a high time variability and associated the source with a supergiant fast X-ray transient. 

In our observation, which was performed one year after the \emph{Suzaku} one, the source is not detected and we only observe a point-like hard source, about $40''$ away, ($\alpha_{J2000}=18^h51^m11.3^s$, $\delta_{J2000}-00^{\circ}00'14.4"$) which has an unabsorbed flux $F\sim1.3\times10^{-13}$ erg cm$^{-2}$ s$^{-1}$.
We conclude that the variable source was in its quiescent state during the \emph{XMM-Newton} observation presented here.

\section{Discussion and conclusions}
\label{Conclusions}

Our analysis of the \emph{XMM-Newton} EPIC observations of Kes~78 revealed in detail the complex morphology of the remnant, characterized by irregular structures mainly elongated along the north-south direction.

Both the image analysis and the spectral analysis show that the X-ray emission of the remnant is soft, being concentrated below 2 keV. We performed a thorough search for a hard X-ray emitting component observed with \emph{Suzaku} by B16. The non-detection of this component in our data clearly shows that the hard emission (which, in the \emph{Suzaku} data, is contaminated by unresolved point-like sources, which we revealed with \emph{XMM-Newton}) cannot be concentrated in narrow rims and this disfavours a synchrotron origin. The background of our EPIC data is relatively high, with a surface brightness $S_h\sim 6\times 10^{-14}$ erg~s$^{-1}$~cm$^{-2}$~arcmin$^{-2}$ in the $2-10$ keV energy band. Therefore, we cannot exclude the presence of a more diffuse, possibly thermal, hard X-ray emission and deeper observation are necessary to address this issue.

We also studied the spatial variations of the physical properties of the X-ray emitting plasma in Kes 78. We found that there are no significant inhomogeneities between the northern and the southern parts of the remnant, in agreement with the findings of B16. However, our analysis shows that the physical conditions are clearly different in region East (which was not completely covered by the \emph{Suzaku} observations) which is characterized by lower values of plasma temperature and higher values of plasma density and $N_H$. All these three aspects clearly indicate that, in the eastern edge of Kes 78, the shock front is propagating in a dense environment.  
Considering that the shock velocity, $v_s$ decreases with increasing ambient density as $1/\sqrt{n}$ and that the post-shock temperature increases as $v_s^2$, we can calculate the expected trend of $EM$ vs. $kT$ ($EM\propto \Lambda/(kT)^2$). This trend is shown by the dotted curve in Fig. \ref{fig:steppar} (calculated by assuming the same $\Lambda$ in all the regions) and is in qualitative agreement with the results of the spatially resolved spectral analysis, though suggesting that the value of $\Lambda$ may be somehow higher in region East than in region Northwest and Southwest.

Our X-ray data analysis therefore, clearly shows that in region East the shock is slowed down by the interaction with a dense medium. This is in very good agreement with the findings of \citet{zc11}, who found indications of a shock-cloud interaction in the East region by analyzing the $^{13}$CO line emission. Left panel of Fig. \ref{fig:COHa} shows the CO and X-ray emission of Kes 78 and confirms that a large molecular cloud is indeed at the position of region East. Also, the H$_\alpha$ emission in the environment of Kes 78 clearly reveals bright filaments in the east, suggestive of a denser environment (right panel of Fig. \ref{fig:COHa}).

Interestingly, the position of the eastern cloud corresponds to the centroid of the diffuse TeV emission observed by $HESS$ (HESS J1852-000, \citealt{kca11}), which is indicated by the yellow cross in Fig. 1. Also, the lack of X-ray synchrotron emission in Kes 78 excludes the presence of TeV electrons and rules out an IC origin for the TeV emission.

We then conclude that the indications of shock-cloud interactions obtained with our image and spectral analysis, together with the non-detection of X-ray synchrotron emitting filaments indicate a hadronic origin for the $\gamma-$ray emission of Kes 78.

\begin{acknowledgements}
We thank the anonymous referee for comments and suggestions. This paper was partially funded by the PRIN INAF 2014 grant “Filling the gap between supernova explosions and their remnants through magnetohydrodynamic modeling and high performance computing”. 
PZ and YC acknowledge support by the NSFC grants 1503008, 11233001 and 11633007, the 973 Program grant 2015CB857100, and grant 20120091110048 from the Educational Ministry of China. 
SSH acknowledges support by the Natural Sciences and Engineering Research Council of Canada and the Canadian Space Agency
\end{acknowledgements}

 \bibliographystyle{aa}

\end{document}